\documentclass[prl,twocolumn,showpacs,floatfix,amsmath,amssymb,superscriptaddress]{revtex4-2}
\usepackage{amsfonts}
\usepackage{stmaryrd}
\usepackage{bbm}
\usepackage{mathrsfs}
\usepackage{tipa}
\usepackage{amssymb}
\usepackage{txfonts}
\usepackage{graphicx}
\usepackage{dcolumn}
\usepackage{epstopdf}
\usepackage[colorlinks,linkcolor=blue,urlcolor=blue,citecolor=blue]{hyperref}
\usepackage{multirow}
\usepackage{subfigure}
\usepackage{url}
\usepackage{upgreek}
\usepackage[utf8]{inputenc}
\usepackage[english]{babel}

\begin{document}

\newcommand*{\cm}{cm$^{-1}$\,}
\newcommand*{\Tc}{T$_c$\,}


\title{Observation of an emergent energy scale close to dimensional reduction \\in a quasi-two-dimensional quantum magnet}

\author{Anneke~Reinold}
\affiliation{Department of Physics, TU Dortmund University, 44227 Dortmund, Germany}

\author{Laur Peedu}
\author{Kirill~Amelin}
\author{Urmas~Nagel}
\author{Toomas~Rõõm}
\affiliation{National Institute of Chemical Physics and Biophysics, 12618 Tallinn, Estonia}

\author{Sven~Luther}
\author{Hannes~Kühne}
\affiliation{Hochfeld-Magnetlabor Dresden (HLD-EMFL), Helmholtz-Zentrum Dresden-Rossendorf, 01328 Dresden, Germany}

\author{Dirk~Wulferding}
\affiliation{Department of Physics and Astronomy, Sejong University, Seoul 05006, Republic of Korea}

\author{Kingshuk~Mukhuti}
\author{Maarten W. de Dreu}
\author{Peter~C. M. Christianen}
\affiliation{HFML-FELIX, Toernooiveld 7, 6525ED Nijmegen, the Netherlands; Institute for Molecules and Materials, Radboud University, Heyendaalseweg 135, 6525, AJ, Nijmegen, the Netherlands}


\author{Dennis~Kudlacik}
\affiliation{Department of Physics, TU Dortmund University, 44227 Dortmund, Germany}
\author{Dmitri Yakovlev}
\affiliation{Department of Physics, TU Dortmund University, 44227 Dortmund, Germany}

\author{Zhiying~Zhao}
\affiliation{
State Key Laboratory of Functional Crystals and Devices, Fujian Institute of Research on the Structure of Matter, Chinese Academy of Sciences, Fuzhou 350108, China}

\author{Taro Nakajima}
\author{Yoshimitsu Kohama}
\affiliation{Institute for Solid State Physics, University of Tokyo, Kashiwa, Chiba 277-8581, Japan}

\author{Thomas~Lorenz}
\affiliation{Institute of Physics II, University of Cologne, 50937 Cologne, Germany}

\author{Franco~Lisandrini}
\author{Corinna Kollath}
\affiliation{Physikalisches Institut, University of Bonn, Nussallee 12, 53115 Bonn, Germany}

\author{Marcin Raczkowski}
\affiliation{Institut für Theoretische Physik und Astrophysik, Universität Würzburg, 97074 Würzburg, Germany}

\author{Fakher F. Assaad}
\affiliation{Institut für Theoretische Physik und Astrophysik, Universität Würzburg, 97074 Würzburg, Germany}
\affiliation{Würzburg-Dresden Cluster of Excellence ctd.qmat, Am Hubland, D-97074 Würzburg, Germany}

\author{Zhe Wang}
\affiliation{Department of Physics, TU Dortmund University, 44227 Dortmund, Germany}

\date{\today}

\begin{abstract}
By appropriately perturbing a critical transverse-field Ising chain away from its critical point, the system can develop a finite correlation length with a characteristic purely massive spectrum, whose ratios and correlations are precisely described by an integrable field theory and an infinite set of integrals of motion corresponding to the $E_8$ Lie algebra.
In this work, we report on experimental observation of a characteristic massive spectrum close to transverse field-induced dimensional reduction in a quasi-two-dimensional quantum magnet Cu$_2$(OH)$_3$Br, providing evidence for an emergent $E_8$ symmetry and the corresponding excitations of bound states in the sublattice of its ferromagnetic chains.
These results demonstrate the power of integrable field theory in describing emergent many-body quantum critical phenomena in condensed matter systems.
\end{abstract}

\maketitle

Exotic quantum many-body dynamics may emerge around a quantum critical point with characteristic dynamical response governed by an emergent symmetry. 
One representative scenario has been frequently discussed in two dimensions:  
In a spin-1/2 square-lattice quantum antiferromagnet,  
a quantum phase transition can occur between a Néel-ordered phase and a valence bond solid \cite{Senthil04,Sandvik07,Jiang_2008,MelkoKaul08,Sandvik10,Singh10,
NahumPRX2015,WangSenthilPRX17,Liu2019}.
The former is characterized by antiferromagnetically ordered spins which break SU(2) symmetry in spin space, while in the latter state spins are paired in singlet valence bonds thereby breaking lattice translational symmetry.
At the quantum critical point these broken symmetries are restored which is characterized by the emergence of deconfined fractional spinon excitations that carry spin-1/2, in contrast to the spin-1 excitations of confined spinons on either side of the quantum critical point.
Such a so-called deconfined quantum critical point, corresponding to a continuous transition between two phases that spontaneously breaks very different symmetries, is beyond the Landau-Ginzburg-Wilson paradigm of phase transitions~\cite{Singh10,Senthil23Review,Mila23,Cui2025}.

A very different scenario can be realized in other models, where characteristic low-energy excitations (such as solitons) emerge not at the quantum critical point, but away from it \cite{EguchiYang89,MussardoPhysRep1992,Mussardo,Lencses2023}.
A paradigmatic model is the spin-1/2 transverse-field Ising chain:
While a quantum critical point induced by the applied transverse field is characterized by massless excitations \cite{Sachdev}, an additional relevant perturbation (e.g. a small longitudinal field) shifts the system away from the quantum critical point, which leads to the emergence of a characteristic excitation spectrum of eight massive particles corresponding to an emergent $E_8$ symmetry in an exactly solvable field theory ($E_8$ affine Toda field theory) with an infinite number of integrals of motion \cite{Zamolodchikov89,Delfino95,Delfino96,WangWu2021}.
Without the relevant perturbation, the $E_8$ symmetry, however, does not exist at or on either side of the one-dimensional transverse field Ising quantum critical point.
Therefore, in both of these two scenarios (either at or off a quantum critical point), an emergent symmetry is manifested by a corresponding emergent characteristic dynamics.

Emergent symmetries and their corresponding characteristic dynamics have been a compelling subject of intensive theoretical studies (e.g. emergent SO(5), O(4) or O(2) symmetries at deconfined quantum critical points  \cite{Senthil04,Sandvik07,Sandvik10,NahumSomoza15,SernaNahum19,MaYouMeng19,
HuangYouXiang19,SatoFakher23,LiuGu24}, or emergent $E_8$, $E_7$ or $E_6$ symmetries away from a quantum critical point \cite{EguchiYang89,MussardoPhysRep1992,Mussardo_2024,FitosTakacs24}).
However, experimental realizations of these theoretical models remain scarce: While a deconfined quantum critical point is possibly realized in SrCu$_2$(BO$_3$)$_2$ under pressure \cite{CuiYu2023,Mila23}, an emergent $E_8$ symmetry was evidenced in the Ising-chain ferromagnet CoNb$_2$O$_6$ \cite{Coldea10,Kjall11,Amelin20,Oshikawa20,Amelin2022} and antiferromagnet BaCo$_2$V$_2$O$_8$ \cite{Zhang20,Zou21,Oshikawa20,Amelin2022,WangLake24} in a transverse magnetic field, away but close to a quantum critical point.
For a large number of other theoretical models, solid-state realizations are most often far from obvious.

In this work, we observe spectroscopic evidence for an emergent energy scale of quantum spin dynamics in a very different model system, i.e. close to a field-induced dimensional-reduction phase transition in a quasi-two-dimensional spin-1/2 quantum magnet Cu$_2$(OH)$_3$Br.
The observed emergent energy scale corresponds to the emergent $E_8$ symmetry that was predicted for the perturbed transverse-field Ising-chain system. 
By performing nuclear magnetic resonance (NMR), Raman, and terahertz (THz) spectroscopic measurements at low temperatures and in high magnetic fields, we investigate evolution of the dynamical responses crossing the phase transition due to field-induced dimensional reduction.

For the $^{81}$Br NMR experiments, we used a commercial phase-coherent spectrometer and a 16 T superconducting magnet. At a given magnetic field, we consistently employed either a 3He sample-in-liquid cryostat or a 4He flow cryostat, with exception of the measurements at 16~T, where both cryostats were used with an overlapping temperature range of the NMR data.
We measured the $^{81}$Br NMR spectra and nuclear spin-lattice
relaxation time $T_1$ by using a standard Hahn spin-echo
pulse sequence and an inversion-recovery method, respectively.
The spectra were recorded by sweeping frequency at fixed magnetic fields.
A single-axis goniometer was used to align several stacked single crystals
of Cu$_2$(OH)$_3$Br.
The high-field Raman measurements were carried out in Voigt geometry at the Nijmegen High Field Magnet Laboratory (HFML-FELIX) using a resistive magnet and a home-built Raman probe inside a $^4$He-bath cryostat. 
The Raman-probe was equipped with a resistive heater element to allow for in situ temperature control. A $\lambda$ = 532 nm laser (Spectra-Physics Millennia Pro) was focused onto the sample with a laser power below 0.13 mW and a spot diameter of about 2 $\mu$m. A volume Bragg grating filter set was used to remove elastically scattered light. The Raman-scattered light was dispersed through a single-stage monochromator (Princeton Instruments HRS-500) onto a liquid-nitrogen cooled charge-coupled device detector (PyLoN eXcelon).
We performed THz absorption measurements of a Cu$_2$(OH)$_3$Br single crystal installed in a cryostat equipped with a superconducting magnet for applying magnetic fields up to 17~T. The sample temperature was maintained at 3~K.
The measurements were performed with a Martin-Puplett interferometer coupled to a $^3$He-cooled Si-bolometer \cite{Wang17,Zhang20,Amelin2022,Pilch23,Pilch25}.
The direction of linear polarization of the THz beam was tunable with
a rotatable polarizer.

\begin{figure}[t]
\centering
\includegraphics[width=1\linewidth]{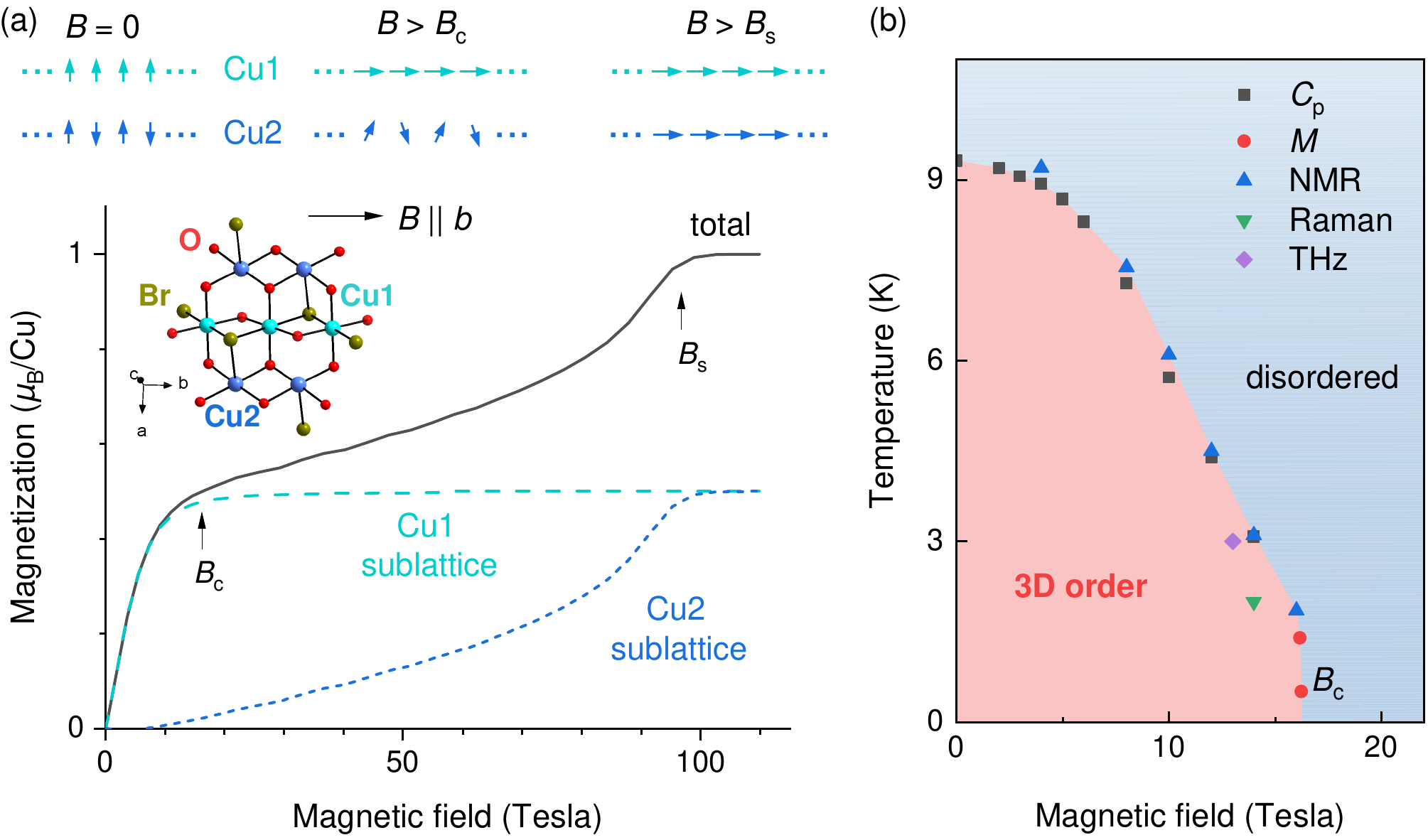}
\caption{
(a) Total and sublattice magnetizations in the direction of the applied magnetic field $B \parallel b$ at zero temperature by quantum Monte Carlo simulations \cite{Reinold25}. A critical field $B_c \simeq 16.3$~T and a saturation field $B_s \simeq 95$~T correspond to the field-induced saturation spin alignment of the Cu1 and the Cu2 sublattices, respectively.
(b) Phase boundary $T_N(B)$ of a three-dimensionally ordered and a field-induced disordered phase determined by specific heat $C_p$, magnetization $M$, NMR $1/T_1$ (Fig.~\ref{fig:NMR}), Raman (Fig.~\ref{fig:Raman}), and THz spectroscopic measurements (Fig.~\ref{fig:IR}).}
\label{fig:Magn_PhaseDia}
\end{figure}

The magnetic structure of Cu$_2$(OH)$_3$Br can be viewed as being constituted by ferromagnetic Cu1 spin chains and antiferromagnetic Cu2 spin chains that are running along the crystallographic $b$ axis and alternately coupled in the $a$ direction [see inset of Fig.~\ref{fig:Magn_PhaseDia}(a)].
The interlayer exchange couplings along the $c$ axis are much weaker than the couplings in the $ab$ plane \cite{ZhaoHe19}. 
At zero field a long-range magnetic order occurring below $T_N=9.3$~K is characterized by a ferromagnetic alignment of the Cu1 spins which are in the \textit{ac} plane and span an angle of 45° from the \textit{a} axis, and antiferromagnetically ordered Cu2 spins which are either parallel or antiparallel to the \textit{a} axis \cite{ZhaoHe19,Reinold25}.
The magnetic structure indicates an effective Ising-type exchange anisotropy, since single-ion anisotropy of Cu$^{2+}$ ions is usually weak.
An Ising exchange anisotropy has been considered to model the spin-wave excitations without external magnetic field~\cite{ZhangKe20}. 
An applied magnetic field along the \textit{b} axis is perpendicular to the ordered spins at zero field, which for the ferromagnetic Cu1 spin chains realizes a similar setting as in a transverse-field Ising chain.

\begin{figure}[t]
\centering
\includegraphics[width=0.9\linewidth]{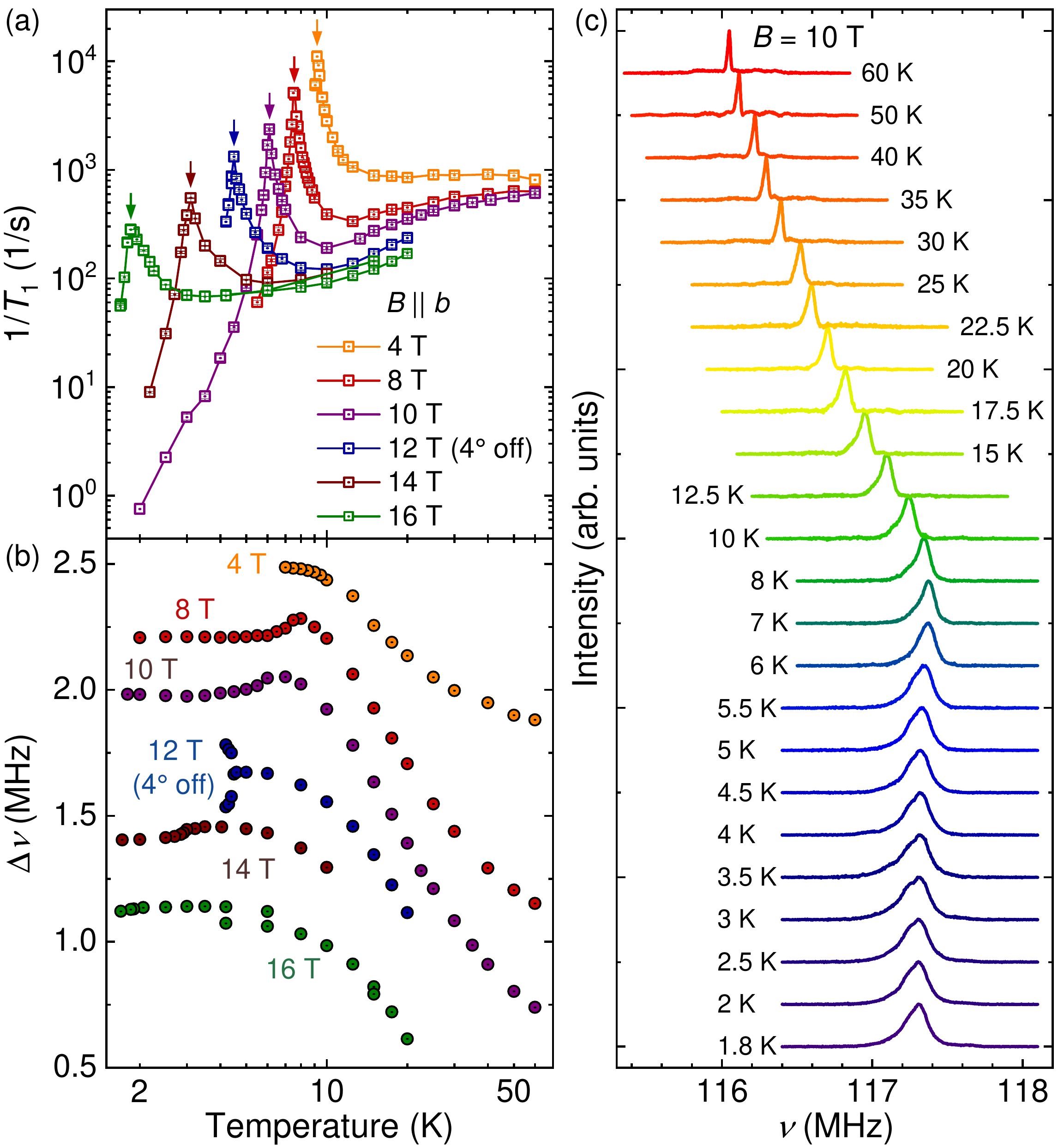}
\caption{
(a)  Temperature-dependent $1/T_1$ rate at different
magnetic fields, applied parallel to the crystallographic $b$ axis. The vertical arrows mark the transition
temperature $T_N$ into the magnetically ordered phase.
(b) Temperature dependence of the $^{81}$Br NMR frequency
shift $\Delta \nu$ for fields between 4 and 16~T.
For clarity, the curves from 10 to 16 T are vertically shifted with respect to each other by $-0.3$~MHz.
(c) Waterfall diagram of frequency-swept spectra at 10~T and temperatures between
1.8 and 60~K.
}
\label{fig:NMR}
\end{figure}

The magnetization process in this transverse field is characterized by two steps, i.e. field-induced polarization of the Cu1 spins above $B_c$ and of the Cu2 spins above $B_s$, respectively. 
As shown by our experimental measurements and theoretical simulations [see Fig.~\ref{fig:Magn_PhaseDia}(a)] \cite{Reinold25}, an initial increase of magnetization is due to a gradual alignment of the Cu1 spins along the applied field direction.
This reduces the Cu1 spin components that are perpendicular to the field, resulting in a decrease of the interchain coupling between the Cu2 and Cu1 spins, since the Cu2 spins remain largely perpendicular to the field. 
Above $B_c \simeq 16.3$~T, the Cu1 spins are fully polarized by the applied field, while the net magnetization of the Cu2 sublattice remains small.
Hence the Cu1 and Cu2 chains are essentially decoupled due to the enhanced one-dimensional fluctuations, leading to the suppression of the three-dimensional long-range magnetic order and the occurrence of dimensional reduction \cite{Reinold25}.
The field-induced phase transition from the magnetically ordered to disordered phase is confirmed by thermodynamic data~\cite{Reinold25} such as specific heat and magnetization [Fig.~\ref{fig:Magn_PhaseDia}(b)], as well as by NMR measurements as will be presented below (Fig.~\ref{fig:NMR}).
According to our previous quantum Monte Carlo simulations, the Cu2 spins are finally polarized above $B_s \simeq 95$~T in the saturation phase \cite{Reinold25}.

To investigate the characteristic features of dynamical responses crossing the field-induced dimensional reduction, we performed NMR, Raman, and THz spectroscopic measurements for applied magnetic fields along the
crystallographic \textit{b} axis.
Figure~\ref{fig:NMR} presents the NMR results for temperatures between 1.8 and 60 K and at magnetic fields up to 16~T.
The observed $^{81}$Br NMR spectral lines correspond to the central ($I_z = -1/2 \leftrightarrow +1/2$) of the nuclear energy levels. We found no satellite transitions within a broad frequency range of several MHz. 
A representative set of the NMR spectral lines at 10~T
and various temperatures is shown in Fig.~\ref{fig:NMR}(c), which exhibits 
a spectral broadening and increasing shift $\Delta\nu$ with decreasing
temperature in the paramagnetic regime.
The obtained parameters $1/T_1$ and $\Delta\nu=\nu_\text{res}-\nu_0$ by fitting the $^{81}$Br NMR spectra are
summarized in Figs.~\ref{fig:NMR}(a) and \ref{fig:NMR}(b), respectively.
$\nu_\text{res}$ denotes the numerical central frequency of the $^{81}$Br NMR spectra.
The reference frequency $\nu_0$ is given by $\nu_{0} = ^{81}\gamma_{n} B_{0}$, where $^{81}\gamma_{n}$ = 11.4989~MHz/T is the nuclear gyromagnetic ratio of $^{81}$Br and $B_{0}$ is the external magnetic field.
Below the phase transition temperature into the magnetically ordered state,
the spectral line is broadened and the shift decreases
slightly, finally becoming constant at very low temperatures.
At 12~T, the sample was intentionally oriented by four
degrees away from $B \parallel b$, revealing a splitting of the spectral
line below $T_N$. This line splitting is a measure of
the local staggered magnetization, and clearly indicates
commensurate long-range antiferromagnetic order. 
For a perfect alignment $B \parallel b$, as adjusted for all other fields, the local staggered magnetization at the $^{81}$Br
positions is effectively zero due to form-factor effects, stemming from the symmetric positioning of the $^{81}$Br nuclei between the neighboring Cu1 and Cu2 spins, so that only the uniform component of the local magnetization is probed below $T_N$.

\begin{figure}[t]
\centering
\includegraphics[width=1\linewidth]{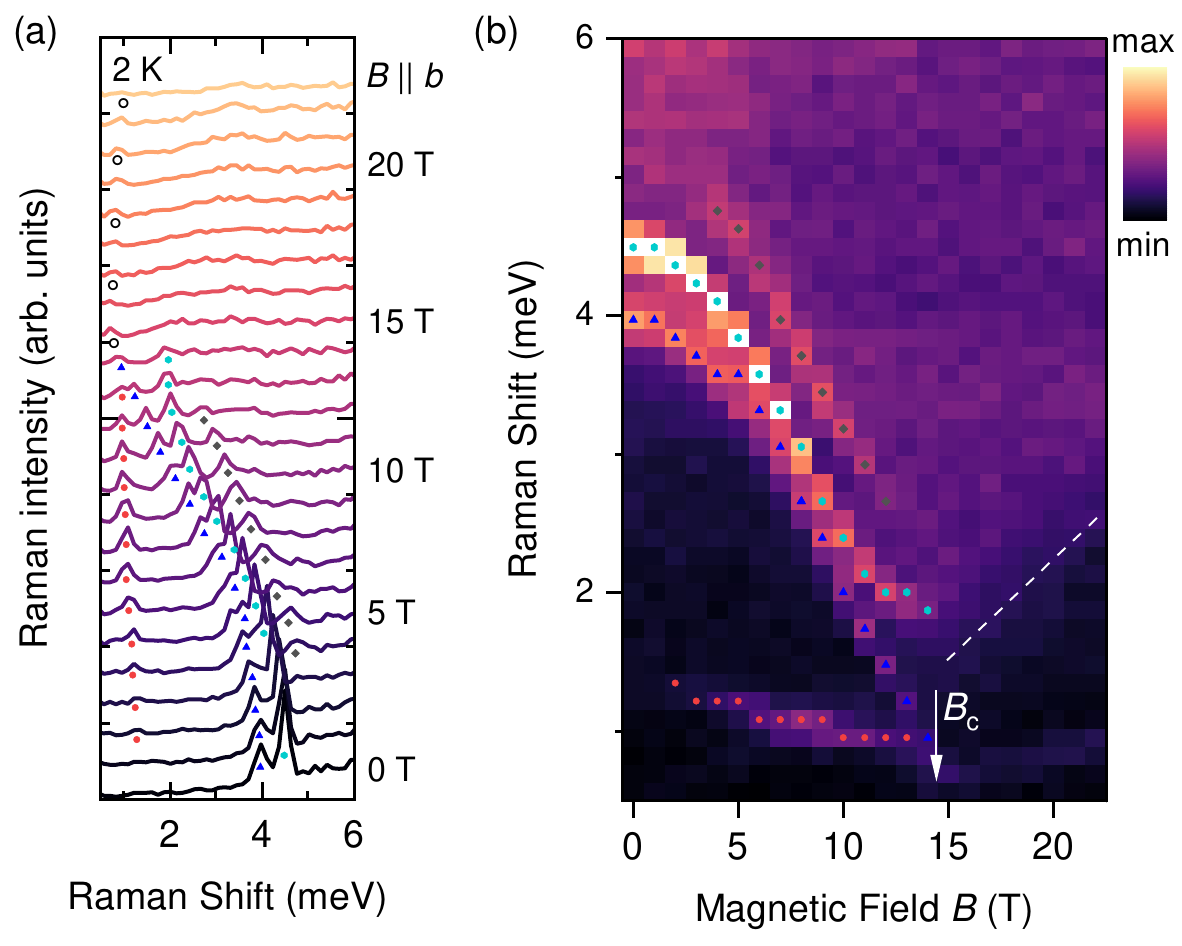}
\caption{
(a) Raman spectra measured at 2~K in various applied magnetic fields $B \parallel b$.
(b) Contour plot of the Raman spectra in (a). The arrow marks the critical field $B_c$. The dashed line is a guide to the eye, indicating shift of the lower boundary of an excitation continuum above $B_c$.}
\label{fig:Raman}
\end{figure}

The spin-lattice relaxation rate $1/T_1$ probes the low-energy
excitations of the electronic spin lattice. The temperature-dependent $1/T_1$ data, recorded at fields between 4 and 16~T, yields sharp maxima, which mark the corresponding transition temperatures $T_N$ into the magnetically ordered phase [Fig.~\ref{fig:NMR}(a)].
These transition temperatures are summarized in the phase diagram in Fig.~\ref{fig:Magn_PhaseDia}(c), and are consistent with the measurements of thermodynamic quantities \cite{Reinold25}.
Below $T_N$, a strong decrease of $1/T_1$ signals the opening of the magnetic excitation gap in the ordered state.

\begin{figure*}[t]
\centering
\includegraphics[width=0.96\linewidth]{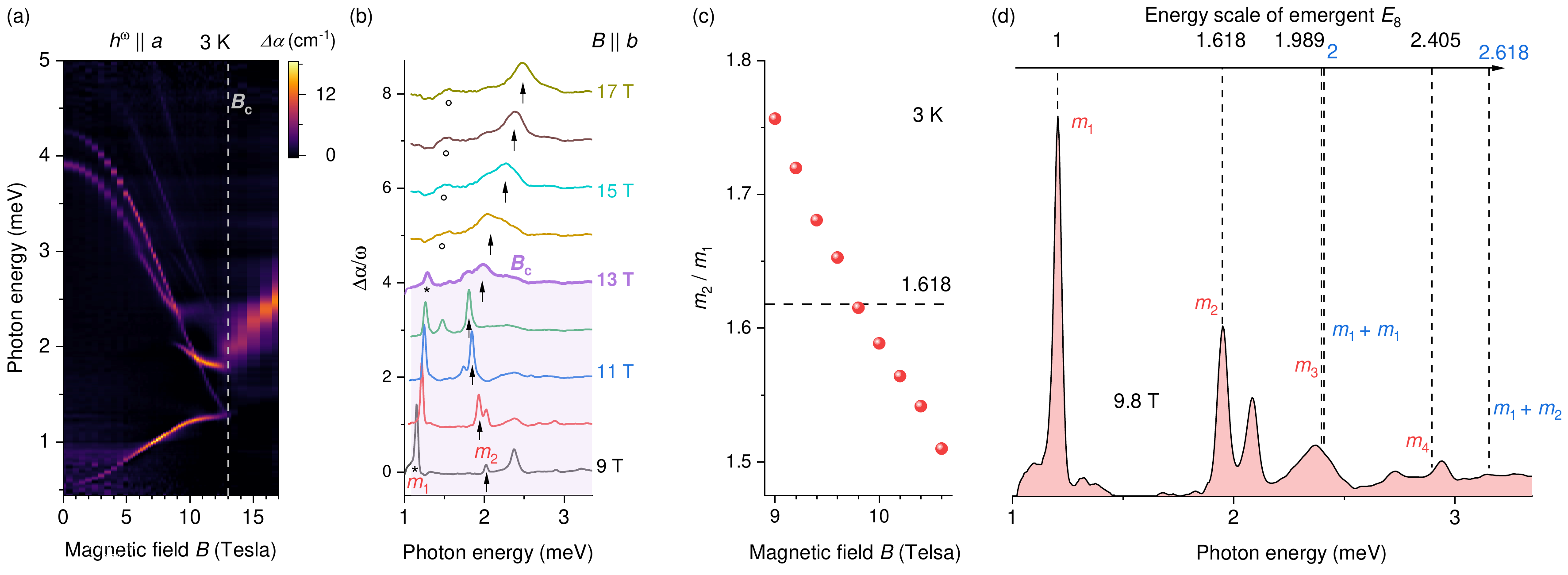}
\caption{
(a) Color-coded plot of THz absorption coefficient measured at 3~K with polarization $h^\omega \parallel a$ for various applied magnetic fields $B \parallel b$ up to 17~T.
(b) THz absorption spectra $\Delta \alpha/\omega$ around the field-induced phase transition at $B_c=13$~T [see Fig.~\ref{fig:Magn_PhaseDia}(b)].
$\Delta \alpha$ denotes field-induced change of absorption coefficient and $\omega$ for wavenumber, both in the unit of cm$^{-1}$.
At the 9~T spectrum the asterisk and arrow indicate the two lowest-lying $m_1$ and $m_2$ modes, respectively.  
(c) Ratio of the $m_2$ and $m_1$ frequencies versus the applied magnetic field. $m_2/m_1$ reaches the golden ratio of 1.618 at $B=9.8$~T, marked by the dashed line. 
(d) $\Delta \alpha/\omega$ spectrum at 9.8~T in comparison to the energy scale (top scale) of emergent quasiparticles governed by the $E_8$ symmetry.}
\label{fig:IR}
\end{figure*}

The gapped low-energy spin excitations are also resolved in our Raman spectroscopic measurements.
Figure~\ref{fig:Raman} shows unpolarized Raman spectra taken at $T = 2$~K with magnetic fields applied along the $b$ axis and increasing in steps of 1~T from 0 to 23 T. The same dataset is presented in Fig.~\ref{fig:Raman}(b) as a contour plot. Here, we focus on the low-energy excitations between 0.5 and 6 meV with all resolved modes marked by symbols.
In the zero-field limit, the magnetic sector exhibits a broad, gapped continuum that has a lower limit of about 4 meV, which corresponds to a continuum of spinon excitations arising from spin correlations within the antiferromagnetic chains as previously resolved by inelastic neutron scattering experiments \cite{ZhangKe20}. 
Conversely, the sharp excitations below $B_c = 14$~T (see scatters in Fig.~\ref{fig:Raman}) are assigned to magnon modes due to the long-range order and reduced spin fluctuations. Such coexistence of fundamentally distinct quasiparticles (spinons and magnons) has been reported in related low-dimensional quantum magnets such as Cu$_2$(OH)$_3$NO$_3$~\cite{wulferding-25}, SeCuO$_3$~\cite{testa-21}, and $\alpha$-RuCl$_3$~\cite{wulferding-20,Sahasrabudhe20}.
When applying magnetic fields along the crystallographic $b$ axis, we observe a shift of the spinon continuum towards lower energies together with the softening of the magnon modes [see Fig.~\ref{fig:Raman}(b)].
A field-induced phase transition above 14~T is clearly indicated by a shift of the continuum's lower boundary again towards higher energy [see the dashed line in Fig.~\ref{fig:Raman}(b)], as well as by the disappearance of the sharp modes (which either disappear or become too broad to be resolved) [Fig.~\ref{fig:Raman}(a)].
These behaviors are signatures of enhanced one-dimensional fluctuations and decoupling of the spin chains in the high-field phase, as also indicated by the two-step magnetization process [see Fig.~\ref{fig:Magn_PhaseDia}(a)] \cite{Reinold25}.

The simultaneous occurrence of dimensional reduction, phase transition, and saturation of the Cu1 spin chains at the applied transverse field $B_c$ is phenomenologically reminiscent of transverse-field Ising-chain quantum phase transition in the quasi-one-dimensional compounds CoNb$_2$O$_6$ \cite{Coldea10,Amelin20,Oshikawa20,Amelin2022} and BaCo$_2$V$_2$O$_8$ \cite{Zhang20,Zou21,Oshikawa20,Amelin2022}, 
where an emergent energy scale of quantum spin excitations corresponding to the $E_8$ symmetry can be observed in the ordered phase below but close to the transverse field-induced phase transition.
To investigate whether $E_8$ symmetry may emerge close to the dimensional reduction in Cu$_2$(OH)$_3$Br, we carried out THz spectroscopic measurements in applied transverse field $B \parallel b$. 

Figure~\ref{fig:IR}(a) summarizes our THz spectroscopic measurements of absorption coefficient in various external fields up to 17~T and for THz magnetic-field polarization $h^\omega \parallel a$. These measurements were performed at 3~K in Voigt transmission geometry with the propagation direction of the THz waves perpendicular to the applied field.
Very rich features of quantum spin excitations are observed from Fig.~\ref{fig:IR}(a).
Whereas in the high-field and zero-field limits only one or two modes are resolved corresponding to the one-dimensional spin excitations \cite{ZhangKe20}, several additional excitation modes appear around 9--10~T.
While the other details will be discussed in a separate work, in Fig.~\ref{fig:IR}(b) we highlight the field-dependent evolution around the phase transition at 13~T corresponding to the sample temperature of 3~K [see also Fig.~\ref{fig:Magn_PhaseDia}(b)].

At 9~T, the spectrum exhibits three modes, with the two lowest-energy ones denoted by $m_1$ and $m_2$, respectively.
Evolution of these two modes with increasing fields can be traced reliably. As indicated by the arrows, at and above $B_c=13$~T the spectra are characterized by a broader excitation mode, which hardens in higher fields in contrast to the softening of the $m_2$ mode below 13~T, indicating a gap opening above $B_c$. These observations clearly indicate a field-induced phase transition at 13~T, in agreement with the thermodynamic measurements [Fig.~\ref{fig:Magn_PhaseDia}(b)] and with Raman and NMR spectroscopic measurements. In the high-field spectra above 13~T, we tentatively mark a lowest-energy broad feature [see circles ("$\circ$") in Fig.~\ref{fig:IR}(b)], which might be an excitation mode corresponding to the lowest-energy Raman mode in Fig.~\ref{fig:Raman}(b) but its magnitude hardly exceeds our experimental uncertainties.

We focus on the discussions of the 9--10 T field range, where a rich set of spectral features is observed [see Fig.~\ref{fig:IR}(a)] and the system approaches the dimensional reduction in the magnetically ordered phase. 
In this field range, we have measured the transmission spectra for more field values and determined the field dependence of the $m_1$ and $m_2$ mode frequencies. The frequency (or mass) ratio $m_2/m_1$ is derived and plotted in Fig.~\ref{fig:IR}(c).     
Because the $m_1$ mode hardens and the $m_2$ mode softens, their ratio decreases monotonically with increasing field.
As indicated by the dashed line in Fig.~\ref{fig:IR}(c), $m_2/m_1$ reaches the golden ratio of 1.618 at 9.8~T. This ratio is predicted for the two lowest massive particles corresponding to the emergent $E_8$ symmetry away from a transverse-field Ising-chain quantum critical point by an integrable Toda field theory~\cite{Zamolodchikov89}.

The integrability of the emergent $E_8$ Toda field theory allows a precise calculation of the emergent dynamical spectra \cite{Zamolodchikov89,Delfino95,Delfino96}.
Theoretical calculations of the correlation functions have revealed not only the single particle excitations of $m_1$, $m_2$, ..., $m_8$, but also for their multiparticle channels \cite{Zhang20,Zou21,WangWu2021}.
In particular, it has been shown that although the energies of the higher-energy single particles (e.g. $m_3 = 1.989 m_1$ and $m_4 = 2.405 m_1$) are close to or above the onset energy of the two-particle $m_1+m_1$ excitation continuum, they are sufficiently stable with strong spectral weights that are well distinguishable from the two-particle $m_1+m_1$ continuum \cite{Zhang20,WangWu2021}. Hence these higher-energy single particles should be experimentally resolvable \cite{Zhang20,Amelin20,Zou21,Amelin2022}.
Therefore, corresponding to the golden ratio, we present in Fig.~\ref{fig:IR}(d) the absorption spectrum of Cu$_2$(OH)$_3$Br at 9.8~T to compare the observed higher-energy excitations with the expected $E_8$ energy scale of single- and two-particle excitations.
As indicated by the dashed lines, the energy scale of the observed excitations is in good agreement with the $E_8$ excitations $m_1$, $m_2$, $m_3$, $m_1+m_1$, $m_4$ and possibly also $m_1+m_2$.
The observation of the characteristic energy scale indicates an emergent $E_8$ symmetry close to the dimensional reduction in Cu$_2$(OH)$_3$Br.

The model that reproduces the magnetization data of Fig.~\ref{fig:Magn_PhaseDia}(a) \cite{Reinold25} as well as the zero-field neutron scattering experiments of Ref.~\cite{ZhangKe20} has SU(2) spin symmetry that is reduced to U(1) in the presence of the applied magnetic field. As such it cannot account for the present experimental data. However, fitting neutron scattering spectra
with linear spin-wave theory suggested that the dominant source
of Ising anisotropy lies in the ferromagnetic chain \cite{ZhangKe20}. Thereby, from the symmetry point of view, the ferromagnetic chain in the field regime $B_c < B < B_s$ is an instance of a transverse-field Ising model in the disordered phase corresponding to the fully polarized state depicted in Fig.~\ref{fig:Magn_PhaseDia}(a).
As the field is reduced, this model will undergo an Ising transition to the ordered phase with broken Ising symmetry.
At the critical point the inter-chain coupling is a relevant perturbation that leads to magnetic ordering.
This ordering endows the one-dimensional transverse field Ising model with a
small longitudinal field, corresponding to the deformation that
allows the emergence of the $E_8$ excitation spectra.

To summarize, by using nuclear magnetic resonance, Raman, and terahertz spectroscopies at low temperatures and in high magnetic fields, we have studied dynamical responses around a transverse field-induced phase transition in a quasi-two-dimensional quantum magnet Cu$_2$(OH)$_3$Br, whose two-dimensional magnetic structure is constituted by alternately coupled ferromagnetic Cu1 and antiferromagnetic Cu2 spin chains. The measured dynamical responses consistently exhibit characteristic features at the boundary of the field-induced quantum phase transition, corresponding to the full polarization of the ferromagnetic Cu1 spin chains which results in an effective magnetic decoupling of the Cu1 and Cu2 spin chains. 
This dimensional reduction causes the suppression of the three-dimensional long-range magnetic order. In particular, we observe an emergent energy scale of the quantum spin excitations in the ferromagnetic Cu1 spin chains, when the system is tuned close to the dimensional reduction.
The observed characteristic energy scale corresponds to an emergent $E_8$ symmetry which was predicted by an emergent integrable field theory in a perturbed transverse-field Ising chain off the quantum critical point.

\begin{acknowledgments}
We acknowledge support by the European Research Council (ERC) under the Horizon 2020 research and innovation programme, Grant Agreement No. 950560 (DynaQuanta).
D.W. acknowledges support from the ITRC Program through the IITP, and from the Global Research Development Center (GRDC) Cooperative Hub Program through the NRF Korea, funded by the (MSIT) (Grant Nos. RS-2024-00437191 and RS-2023-00258359).
This work was also supported by HLD-HZDR and HFML-FELIX, members of the European Magnetic Field Laboratory (EMFL).
F.F.A. acknowledges support by the Würzburg-Dresden Cluster of Excellence ctd.qmat (EXC 2147, Project No. 390858490). The numerical simulations
were performed with the ALF-implementation of the auxiliary-field quantum Monte Carlo method \cite{Assaad25}. F.F.A. and M.R. acknowledge the scientific support and HPC resources provided by the Erlangen National High Performance Computing Center (NHR@FAU) of the Friedrich-Alexander-Universität Erlangen-Nürnberg (FAU) under NHR project 80069 provided by federal and Bavarian state authorities.
NHR@FAU hardware is partially funded by the German Research Foundation (DFG) through grant 440719683.
\end{acknowledgments}

\bibliographystyle{apsrev4-2}
\bibliography{COHB_bib}



\end{document}